\documentstyle[epsfig,sprocl]{article}

\def\bg#1{\mbox{\boldmath$#1$}}

\newcommand{\sumint}{\hbox{$\sum$}\!\!\!\!\!\!\int}
\newcommand{\Tr}{\mbox{Tr}}

\newcommand{\del}{\partial}
\newcommand{\beq}{\begin{eqnarray}}
\newcommand{\eeq}{\end{eqnarray}}
\newcommand{\be}{\begin{eqnarray*}}
\newcommand{\ee}{\end{eqnarray*}}

\newcommand{\bx}{{\bf x}}

\newcommand{\ra}{\rightarrow}

\newcommand{\nn}{\nonumber}

\newcommand{\ex}[1]{\langle\,#1\,\rangle}

\def\square{\vcenter{\vbox{\hrule height.4pt
          \hbox{\vrule width.4pt height6pt
          \kern6pt\vrule width.4pt}\hrule height.4pt}}}
\def\boxx{\square}

\begin{document}

\title{EFFECTIVE FIELD THEORY OF QED VACUUM FLUCTUATIONS\footnote{Plenary talk
      at {\it PASCOS-98}, Boston, Massachusetts, USA,  March 22 - 29, 1998.}}
\author{FINN RAVNDAL}
\address{NORDITA, Blegdamsvej 17, DK-2100 Copenhagen  \O \footnote
{Permanent address: Institute of Physics, University of Oslo,   
                                N-0316 Oslo, Norway.}}
\maketitle

\begin{abstract}  Only photons are dynamical degrees of freedom in the QED vacuum 
at energies well below the electron mass $m$. Their interactions via couplings to 
virtual electron-positron pairs are described to lowest order by an effective theory
incorporating the Uehling and Euler-Heisenberg interactions as dominant terms. 
By a redefinition of the electromagnetic field, the Uehling term is shown not to 
contribute in the absence of matter. The Stefan-Boltzmann energy for blackbody 
radiation at temperature $T$ is then modified by a term proportional to $T^8/m^4$. 
Correspondingly, the Casimir force between two parallel plates with separation $L$ 
gets an additional contribution proportional to $1/L^8m^4$. Higher order corrections 
to these results are discussed.
\end{abstract}
  
Before the Standard Model was established, effective field theories were said to
be given by phenomenological Lagrangians. The reason was that they were in general
non-renormalizable in the sense of involving interactions with dimensions
larger than four. As quantum theories they could therefore only give meaningful
results at tree level, i.e. from Feynman diagrams without loops. Besides the Fermi 
Lagrangian for weak interactions, the prototype was the non-linear $\sigma$-model 
for low-energy pion interactions. In the exponential representation it can be 
written in the compact form 
\beq
    {\cal L} = {1\over 4}f_\pi^2\Tr\left[\del_\mu U \del^\mu U^\dagger 
             + m_\pi^2( U + U^\dagger)\right]                            \label{L1}
\eeq
with the $SU(2)$ matrix $U = \exp (i{\bg\tau}\cdot{\bg\pi}/f_\pi)$ where the isospin 
vector ${\bg\pi}= (\pi^+, \pi^0, \pi^-)$ contains the pion fields and $f_\pi$ is 
the pion decay constant. When expanded in powers of the field, one obtains to 
lowest non-trivial order the interacting Lagrangian
\beq
    {\cal L} &=& {1\over 2}\left(\del_\mu{\bg\pi}\cdot\del^\mu{\bg\pi}
             - m_\pi^2{\bg\pi}\cdot{\bg\pi}\right)     \nn \\
    &+& {m_\pi^2\over 24f_\pi^2}({\bg\pi}\cdot{\bg\pi})^2 + {1\over 6f_\pi^2}
     \left[({\bg\pi}\cdot\del_\mu{\bg\pi})^2 
     - ({\bg\pi}\cdot{\bg\pi})(\del_\mu{\bg\pi})^2\right] +\ldots       \label{L2}
\eeq
It was shown by Weinberg that in lowest order perturbation theory these interaction
terms reproduced the scattering lengths obtained from current 
algebra \cite{Weinberg_1}. This was equivalent to using the Lagrangian just as a
classical theory. It took twelve years before it was considered a full-fledged
quantum theory and one-loop corrections were calculated, again by 
Weinberg \cite{Weinberg_2}. A much more detailed investigation  of the theory and
its physical content were then later initiated by Gasser and Leutwyler \cite{GL}. 
Divergences from loop integrations are absorbed by the coupling constants of higher order 
interactions \cite{Ecker}.

One can only speculate why this development took so many years. One reason can be
that the interactions in Eq.(\ref{L2}) involve time-derivatives which were notoriously
difficult to handle before when path integrals were not generally used in field 
theory. Also, divergences must be regularized by a method which preserves
chiral invariance. Today one uses dimensional regularization which was first fully 
developed in the mid-seventies.

A similar and even slower development has taken place in the understanding  of 
quantum effects in Einstein's theory of gravity described by the Hilbert action
\beq
        S[g] = {2\over \kappa^2}\int\! d^4x \sqrt{-g} R
\eeq
with $\kappa^2 = 32\pi G_N$. Here $g$ is the determinant of the metric 
$g_{\mu\nu}(x)$ and $R$ is the scalar curvature given by derivatives of the metric.
Expanding around flat spacetime by writing $g_{\mu\nu} = \eta_{\mu\nu} 
+ \kappa h_{\mu\nu}$ one gets in addition to the free Lagrangian for the graviton
field $h_{\mu\nu}$, an infinite series of interactions of dimensions more than four.
The theory is thus not renormalizable in the textbook sense. But a few years ago,
Donoghue showed that treating the above action as describing an effective theory
for energies below the Planck energy $1/\kappa$, one can systematically calculate
quantum corrections to classical results \cite{Donoghue}. For example, Newton's 
constant $G_N$ becomes effectively smaller at shorter distances which is 
characteristic for a non-abelian gauge theory. These quantum effects are completely 
negligible in practically all realistic physical situations.

Interactions between photons and electrons are described by QED. This is a renormalizable
quantum theory and can thus in principle be used to arbitrarily high energies in the absence
of other particles or interactions. At energies below the electron mass $m$ only photon degrees 
of freedom can be excited from the vacuum and one can then construct an effective but 
non-renormalizable theory for these excitations alone.
Formally this can be done by integrating out the electron field in the full
QED partition function. One can then arrange the result as an expansion in Lorentz and
gauge invariant operators of increasing dimensions. Including operators up to dimension $D=8$
we then have for the effective Lagrangian $ {\cal L}_{eff} = -{1\over 4}F_{\mu\nu}^2 
+ {\cal L}_U + {\cal L}_{EH} + \ldots$ The first term
\beq
     {\cal L}_U = {\alpha\over 60\pi m^2}F_{\mu\nu}\boxx\, F^{\mu\nu}           \label{LU}
\eeq
is the Uehling interaction \cite{Uehling} due to the lowest-order vacuum polarization loop 
where $\alpha = e^2/4\pi$ is the fine structure constant and $\boxx \equiv \del_\mu\del^\mu$. 
The next term
\beq
    {\cal L}_{EH} =  {\alpha^2\over 90m^4}\left[(F_{\mu\nu}F^{\mu\nu})^2 
                    + {7\over 4}(F_{\mu\nu}\tilde{F}^{\mu\nu})^2\right]        \label{LEH}
\eeq
is the lowest order Euler-Heisenberg interaction \cite{EH} where 
$\tilde{F}_{\mu\nu} = {1\over 2}\epsilon_{\mu\nu\rho\sigma}F^{\rho\sigma}$ is the dual field 
strength. 

The equation of motion for the free field is $\boxx\, F_{\mu\nu} = 0$ 
and thus we see that the  Uehling term can be effectively set equal to zero as long as there is 
no matter 
present. Since the equation of motion is only satisfied by on-shell photons, one may  question 
the validity of this simplification where the interactions are used to generate loop diagrams 
with virtual photons. But since one is in general allowed to shift integration variables in the
corresponding functional integrals, we see that under the transformation
\beq
     A_\mu \ra A_\mu + {\alpha\over 30\pi m^2}\boxx A_\mu                      \label{trans}
\eeq
the Uehling interaction is again removed. We are thus left with the result
\beq
    {\cal L} = -{1\over 4}F_{\mu\nu}^2 
                      + {\alpha^2\over 90m^4}\left[(F_{\mu\nu}F^{\mu\nu})^2 
                    + {7\over 4}(F_{\mu\nu}\tilde{F}^{\mu\nu})^2\right]        \label{LEFF}  
\eeq
for the effective theory describing interacting photons at low energies in the absence of matter.

The energy density of free photons at non-zero temperature $T$ is given by the Stefan-Boltzmann 
law ${\cal E} = \pi^2T^4/15$. From the above effective theory we can now calculate the 
first quantum correction to this classical result. 
For dimensional reasons we see that it must thus vary with the temperature like $T^8/m^4$. 
Its magnitude is most directly calculated using the Matsubara formalism where the photon field 
is periodic in imaginary time. In lowest order perturbation theory the correction to the free 
energy density is then obtained directly from the partition function as
$\Delta{\cal F} = \ex{\Delta{\cal L}_E}$ where $\Delta{\cal L}_E$ is the Euclidean version of
the Euler-Heisenberg interaction (\ref{LEH}). We thus have
\beq
   \Delta {\cal F} = {\alpha^2\over 90m^4}\ex{7F_{\mu\nu}F_{\nu\beta}F_{\beta\alpha}F_{\alpha\mu}
   - {5\over 2}F_{\mu\nu}F_{\nu\mu}F_{\beta\alpha}F_{\alpha\beta}}          \label{DF}
\eeq
Expanding the expectation values using Wick's theorem, we see that the result will follow from
the two-loop diagram in Fig.1. This is just a product of two one-loop diagrams and is therefore
much easier to evaluate than the corresponding three-loop digram which would be needed in QED.
\begin{figure}[htb]
 \begin{center}
  \epsfig{figure=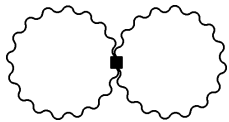,height=15mm}
 \end{center}
 \vspace{-4mm}            
 \caption{Euler-Heisenberg correction to the free energy.}
 \label{fig4}
\end{figure}
For our purpose we need the gauge invariant correlator of the electromagnetic field tensor
\beq
     \ex{F_{\mu\nu}(k)F_{\alpha\beta}(-k)} = {1\over k^2}
     \left[k_\mu k_\beta\delta_{\nu\alpha} - k_\mu k_\alpha\delta_{\nu\beta} 
     + k_\nu k_\alpha\delta_{\mu\beta} -  k_\nu k_\beta\delta_{\mu\alpha}\right]      \label{FF}
\eeq
When the photon field is in thermal equilibrium at temperature $T$, the fourth component of the
four-momentum vector $k_\mu$ is quantized, $k_4 = \omega_n = 2\pi T n$ where the Matsubara index 
$n$ takes all positive and negative integer values for bosons. Loop integrations are then done 
by the sum-integral
\be
     \sumint_k = T \!\!\sum_{n=-\infty}^\infty \!\int{d^3 k\over (2\pi)^3}            
\ee
where the 3-dimensional momentum integration is dimensionally regularized and the Matsubara 
summation is regularized using zeta-functions. In our case Eq.(\ref{DF}) reduces to
\be
    \Delta {\cal F} = - {22\alpha^2\over 45 m^4}\sumint_p \sumint_q \frac{(p\cdot q)^2}{p^2q^2}
\ee
where the basic sum-integral is
\beq
    \sumint_k {k_\mu k_\nu \over k^2}  = {\pi^2 T^4\over 90}                    
    (\delta_{\mu\nu} - 4\delta_{\mu 4}\,\delta_{\nu 4})
\eeq
Many other more complex sum-integrals have been calculated by Arnold and Zhai \cite{AZ} and 
Braaten and Nieto \cite{BN}. We thus obtain \cite{KR}
\beq
    \Delta{\cal F} =  - {22\pi^4\alpha^2 \over 3^5\cdot 5^3} {T^8\over m^4}      \label{DeltaF}
\eeq
This result has previously been derived by Barton \cite{Barton} using a semi-classical method and
treating the interacting photon gas as a material medium. It gives directly the pressure in the
gas as $P = -{\cal F}$. The entropy is given by the derivative with respect to temperature and thus
the energy density follows from ${\cal E} = (1 - T\del/\del T){\cal F}$ as
\beq
    {\cal E} = {\pi^2\over 15}T^4 + {7\cdot 22\pi^4\alpha^2 \over 3^5\cdot 5^3} {T^8\over m^4}  
                         \label{ergdens}
\eeq
Obviously the corrections due to the Euler-Heisenberg interaction are negligible for ordinary 
temperatures. 

Instead of this global derivation of the energy density one can retrieve the same results
from a local calculation based upon the energy-momentum tensor $T_{\mu\nu}(x)$ of the interacting
photon field. In a curved spacetime it is in general defined by
\beq
     T_{\mu\nu}(x) = - {2\over \sqrt{g}}{\delta S[A]\over \delta g^{\mu\nu}(x)}      \label{tmunu}
\eeq
where $S[A]$ is the corresponding action functional for the system and $g$ is the determinant
of the Euclidean metric $g_{\mu\nu}$. The result in flat spacetime now follows from the
effective Lagrangian (\ref{LEFF}). It is given by the standard tensor
\beq
     T_{\mu\nu}^M = F_{\mu\lambda}F_{\nu\lambda} 
                  - {1\over 4} \delta_{\mu\nu}F_{\alpha\beta}F_{\alpha\beta}          \label{TMax}
\eeq
in free Maxwell theory plus a correction 
\beq
     T_{\mu\nu}^{EH} &=& {\alpha^2\over 45 m^4}
    \left(- 4F_{\mu\lambda}F_{\nu\lambda}(F_{\alpha\beta}F_{\alpha\beta})
      +  7F_{\mu\lambda}\tilde{F}_{\nu\lambda}(F_{\alpha\beta}\tilde{F}_{\alpha\beta})\right.\nn \\
     &+& \left.{1\over 2}\delta_{\mu\nu}\left[(F_{\alpha\beta}F_{\alpha\beta})^2 
              - {7\over 4}(F_{\alpha\beta}\tilde{F}_{\alpha\beta})^2\right]\right) \label{TEH}
\eeq
due to the Euler-Heisenberg interaction. The energy density (\ref{ergdens}) can now be obtained
from the expectation value ${\cal E}  = -\ex{T_{44}}$. This approach represents a much more
technically cumbersome approach than the previous method based upon thermodynamics \cite{KR}.

Instead of having the photon field at finite temperature corresponding to periodic boundary
conditions in the imaginary time dimension, we can consider the field confined between two
parallel plates with separation $L \gg 1/m$. The vacuum energy of free photons then give
rise to the Casimir force between the plates. It can be obtained from the fluctuations
of the transverse $(E_x, E_y)$  and longitudinal $E_z$ components of the electric 
field ${\bf E}$ where the $z$-axis is normal to the plates. In terms of the function
\beq
     F(\theta) = -{1\over 2}{d^3\over d\theta^3}\cot{\theta} 
               = {3\over\sin^4{\theta}} - {2\over\sin^2{\theta}}                 \label{F}
\eeq
where $\theta = z\pi/L$ gives the distance from one plate, one has \cite{LR}.
\be
    \ex{{E}_x^2(\bx)} = \ex{{E}_y^2(\bx)}
                 &=& - {\pi^2\over 48L^4}\left({1\over 15} - F(\theta)\right)       \\
    \ex{{E}_z^2(\bx)} &=& + {\pi^2\over 48L^4}\left({1\over 15} + F(\theta)\right)   
\ee
after regularization. Similar expressions obtain for the fluctuations of the magnetic field,
i.e. $\ex{B_x^2} = \ex{B_y^2} = - \ex{E_z^2}$ and $\ex{B_z^2} = - \ex{E_x^2}$. 
The vacuum energy density is then simply ${\cal E} = {1\over 2} \ex{{\bf E}^2(\bx) 
+  {\bf B}^2(\bx)} = - \pi^2/720 L^4$ since the position dependence via the function
$F(\theta)$ drops out in the sum.

Since we know the free photon propagator betwen the plates, it is now possible to calculate
the first quantum correction to the Casimir energy due to the Euler-Heisenberg interaction 
in the effective Lagrangian (\ref{LEFF}). In lowest order perturbation theory it
is given by
\beq
    \Delta E = - \int\!d^3x \;\ex{{\cal L}_{EH}}                              \label{DE}
\eeq
The field correlators diverge near the plates, but these divergences disappear after integration
over the volume between the plates \cite{KR}. One then finds \cite{PRL}
\beq
    \Delta E = - {11\alpha^2\pi^4\over 2^7\cdot 3^5\cdot 5^3 \, m^4L^7}    \label{DE0}
\eeq
Dividing by the plate separation $L$ to get an energy density, we see that it can also be 
obtained from the free energy correction (\ref{DeltaF}) by the substitution $T \ra 1/2L$ just 
as for the leading term of the Casimir energy. It represents a tiny, additional contribution 
to the attractive force between the plates. 

Obviously, the same result for the vacuum energy in the interacting theory should be obtainable 
from full QED. Such a calculation have been attempted by Robaschik and collaborators \cite{Bordag}.
They obtain a correction to the Casimir force varying with the separation as $1/L^5$ while the 
above approach gives a correction going like $1/L^8$. They assume 
that the electrons don't feel the presence of the plates and their result is due to properties of 
the photon propagator at the high energy scale $m$. At such short scales also properties of the
confining plates should in principle enter the result. On the other hand, in the effective 
theory only photon degrees of freedom with much smaller momenta are included. 

From the interacting energy-momentum tensor (\ref{TEH}) one can find the energy density and 
pressure between the plates. This calculation has not yet been completed. It is much more
difficult than in the finite-temperature case since there is no translation invariance
normal to the plates.

The Euler-Heisenberg interaction (\ref{LEH}) was originally derived for constant fields only. 
Here we have used it as an effective Lagrangian for general fields. One might worry that it
then no longer incorporates all the physics. But here the crucial concept of matching comes in.
It is essential in the construction of effective theories. From gauge and Lorentz invariance
we know that the structure of the dimension $D=8$ interaction must be of the form (\ref{LEH})
plus the dimension $D=8$ Uehling interaction. This latter is of the same form as (\ref{LU})
but involving two $\boxx$ operators. It can also be removed by a transformation similar to
(\ref{trans}). The only uncertainty in the $D=8$ operators is then their coefficients. In order 
to find these to lowest order in $\alpha$, we go to the special case of constant fields and use 
the Euler-Heisenberg result. Higher order corrections to the same coefficients have been
calculated by Ritus \cite{Ritus} and confirmed by Reuter, Schmidt and Schubert \cite{RSS}.
Having fixed the coupling constants in this way, we then have the effective Lagrangian 
(\ref{LEFF}) which is valid for arbitrary fields both classically and as a quantum theory. 

Historically, the Euler-Heisenberg interaction was first used by Euler to calculate the 
cross-section for elastic scattering of light by light at energies below the electron mass 
\cite{Euler}. An attempt to calculate the lowest order quantum correction to this result 
was apparently first made by Halter \cite{Halter} who considered the contribution from a 
one-loop diagram with two Euler-Heisenberg four-photon vertices. It was later pointed out 
that this contribution is much smaller than ${\cal O}(\alpha)$ corrections to the coupling 
constants at tree level \cite{Balholm}. At somewhat higher energies a dimension $D=10$ 
interaction comes in and gives the dominant contribution as recently shown by Dicus, Kao and
Repko \cite{DKR}. They find that it can be written as
\be
     {\cal L}_{DKR} &=& {\alpha^2\over 945 m^6}\left[(\del^\alpha\del_\beta F_{\mu\nu})
     F^{\mu\nu}F_{\alpha\lambda}F^{\lambda\beta} 
     + 3(\del^\alpha F_{\mu\nu})(\del_\alpha F^{\mu\nu})F_{\lambda\rho}F^{\lambda\rho}\right. \\
&+&\left. 11\,(\del^\alpha F_{\mu\nu})(\del_\alpha F_{\lambda\rho})F^{\nu\lambda}F^{\rho\mu}\right]
\ee
In addition, they construct a $D=12$ four-photon vertex operator with four derivatives which 
contributes at even higher energies to the scattering amplitude. At this order also the $D=12$ 
Euler-Heisenberg interaction which couples six photons, can contribute by contracting a 
pair of photon fields. However, using dimensional regularization this contribution vanishes 
\cite{DKR}. All these new operators will contribute to the QED vacuum energy when calculating 
to higher orders. We will then also experience that the tree-level coupling constants in the
corresponding effective theory must be renormalized as in ordinary, renormalizable quantum 
field theories.

\end{document}